\documentclass[aps,pra,twocolumn,amsmath,amssymb,nofootinbib,showpacs,superscriptaddress]{revtex4-1}
\usepackage[english]{babel}
\usepackage{latexsym}
\usepackage{graphics}
\usepackage{graphicx}
\usepackage{epsfig}
\usepackage{color}
\usepackage{bm}
\usepackage{amsmath}
\usepackage{amssymb}
\usepackage{amsthm}
\usepackage{dcolumn}
\usepackage{bm}
\usepackage{float}
\usepackage{hyperref}
\usepackage{color}
\usepackage{epstopdf}
\usepackage{cleveref}
\usepackage[svgnames]{xcolor}
\hypersetup{hidelinks,colorlinks=true,allcolors=DarkBlue}

\newcommand{\Tr}{{\rm Tr}\,}

\newcommand{\ket}[1]{|#1\rangle}
\newcommand{\bra}[1]{\langle#1|}

\newcommand{\EindAB}{\mathcal{E}_{A\rightarrow B}^\mathrm{ind}}
\newcommand{\EindBA}{\mathcal{E}_{B\rightarrow A}^\mathrm{ind}}
\newcommand{\EcomAB}{\mathcal{E}_{A\rightarrow B}^\mathrm{com}}
\newcommand{\EcomBA}{\mathcal{E}_{B\rightarrow A}^\mathrm{com}}
\newcommand{\EmixAB}{\mathcal{E}_{A\rightarrow B}^\mathrm{mix}}
\newcommand{\EmixBA}{\mathcal{E}_{B\rightarrow A}^\mathrm{mix}}
\newcommand{\Eind}{\mathcal{E}^\mathrm{ind}}
\newcommand{\Ecom}{\mathcal{E}^\mathrm{com}}
\newcommand{\Emix}{\mathcal{E}^\mathrm{mix}}
\newcommand{\rhoin}{\rho_\mathrm{in}}
\newcommand{\rhom}{\rho_0}
\newcommand{\psitp}{\psi_{\theta,\phi}}

\begin{document}

\preprint{APS/123-QED}

\title{Bidirectional imperfect quantum teleportation with a single Bell state}
\author{E.O. Kiktenko}
\affiliation{Theoretical Department, DEPHAN, Skolkovo, Moscow 143025, Russia}
\affiliation{Bauman Moscow State Technical University, Moscow 105005, Russia}
\affiliation{Geoelectromagnetic Research Centre of Schmidt Institute of Physics of the Earth, Troitsk, Moscow Region 142190, Russia}
\author{A.A. Popov}
\affiliation{Theoretical Department, DEPHAN, Skolkovo, Moscow 143025, Russia}
\affiliation{Bauman Moscow State Technical University, Moscow 105005, Russia}
\author{A.K. Fedorov}
\affiliation{Theoretical Department, DEPHAN, Skolkovo, Moscow 143025, Russia}
\affiliation{Russian Quantum Center, Skolkovo, Moscow 143025, Russia}
\affiliation{LPTMS, CNRS, Univ. Paris-Sud, Universit\'e Paris-Saclay, Orsay 91405, France}

\date{\today}

\begin{abstract}
We present a bidirectional modification of the standard one-qubit teleportation protocol, 
where both Alice and Bob transfer noisy versions of their qubit states to each other by using single Bell state and auxiliary (trigger) qubits.
Three schemes are considered: 
the first, where the actions of parties are governed by two independent quantum random triggers, 
the second with single random trigger, 
and the third as a mixture of the first two.
We calculate the fidelities of teleportation for all schemes and find a condition on correlation between trigger qubits in the mixed scheme 
which allows us to overcome the classical fidelity boundary of 2/3.
We apply the Choi-Jamiolkowski isomorphism to the quantum channels obtained in order to investigate an interplay between their ability to transfer the information, 
entanglement breaking property, and auxiliary classical communication needed to form correlations between trigger qubits.
The suggested scheme for bidirectional teleportation can be realized by using current experimental tools. 

\begin{description}
\item[PACS numbers]
03.65.Wj, 03.65.-w, 03.67.-a
\end{description}
\end{abstract}

\maketitle

\section{Introduction}

Significant progress of experimental techniques for the synthesis and manipulation of individual quantum objects \cite{Lukin}
is a milestone on the path toward
the implementation of a new generation of secure communications with quantum key distribution~\cite{Gisin0}, 
ultrasensitive metrological devices based on quantum sensing~\cite{Ye}, 
and high performance quantum information processors~\cite{Ladd}.

Quantum teleportation \cite{Wootters} is of paramount importance for quantum science and technologies \cite{Pirandola}.
It can be used as a key building block for quantum information technologies such as long-distance quantum communications \cite{Zoller,Gisin2,Kimble,Lukin2} and universal computing \cite{Chuang}.
Due to this fact, quantum teleportation has attracted a great deal of interest in experiments with various substrates such as photonic (both discrete and continuous-variables) qubits 
\cite{Zeilinger,Popescu,Furusawa,Kulik,Gisin3,Bowen,Zhang,Furusawa2,Pan,Walmsley,Walmsley},
nuclear magnetic resonance \cite{Nielsen2},
trapped atoms~\cite{Blatt,Wineland,Riebe,Olmschenk,Nolleke},
atomic ensembles~\cite{Polzik,Polzik2,Chen},
and solid state~\cite{Wallraff,Gao,Bussieres,Pfaff}.

\begin{figure}
\includegraphics[width=0.475\textwidth]{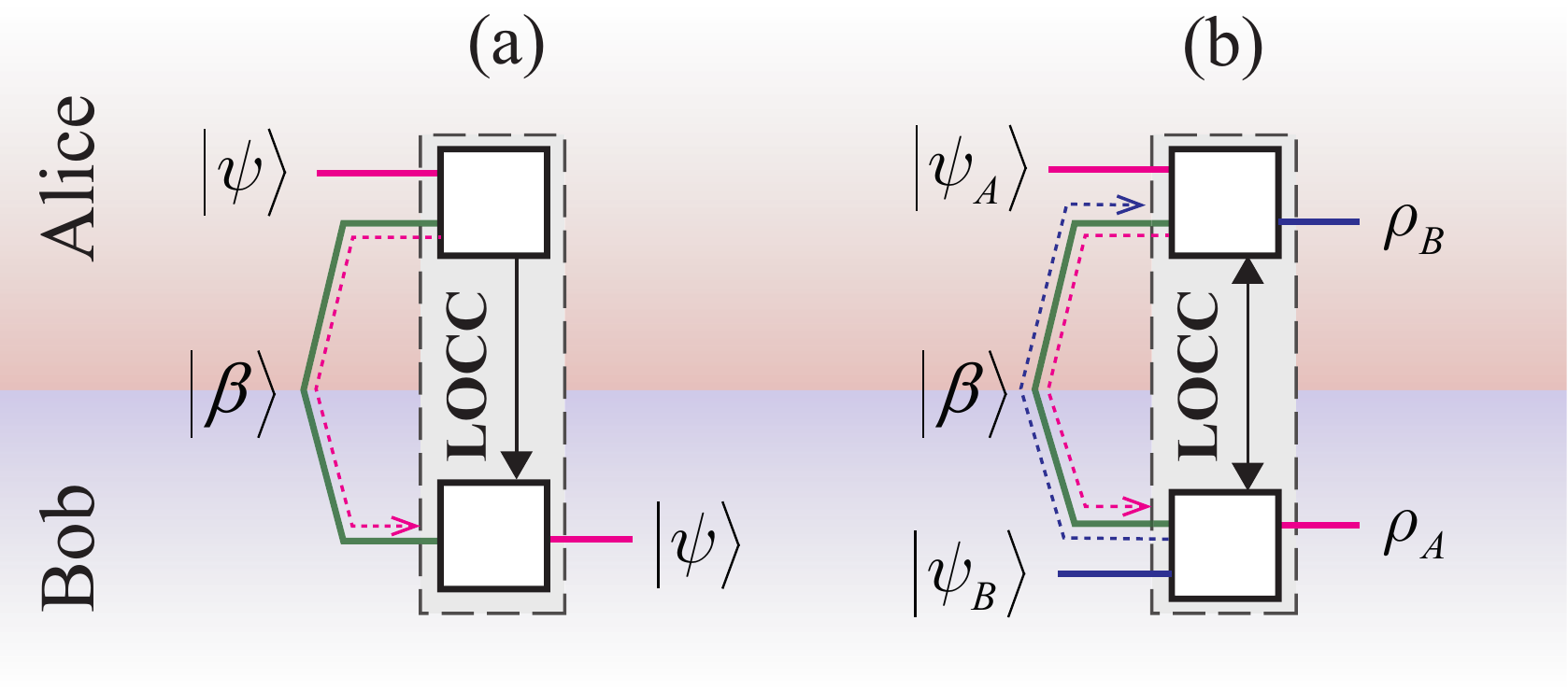}
\vskip -4mm
\caption{Quantum teleportation protocol with local operations and classical communication (LOCC). 
(a) Standard scheme of perfect unidirectional quantum teleportation: 
the qubit state $\ket{\psi}$ is perfectly teleported from Alice to Bob via the shared Bell state $\ket{\beta}$.  
(b) Suggested scheme of imperfect bidirectional quantum teleportation: 
Alice and Bob simultaneously transmit their states $\ket{\psi_A}$ and $\ket{\psi_B}$ to each other by using the Bell state $\ket{\beta}$, 
but they receive the states with density matrices $\rho_A$ and $\rho_B$, which are not perfect versions of the transferred states.}
\label{fig:F0}
\end{figure}

The key ingredient of the seminal teleportation protocol \cite{Wootters} is sharing of a two-qubit maximally entangled Bell state.
Together with local operations and classical communication (LOCC) it allows the establishment of an ideal one-qubit channel from Alice to Bob [see Fig.~\ref{fig:F0}(a)].
This process can be interpreted as a formation of an information-transferring medium for a qubit state by means of the Bell state \cite{Penrose,Laflamme,Hardy},
which is very attractive from the viewpoint of the investigation of fundamental aspects of quantum physics~\cite{EPR}. 
From the viewpoint of such treatment, quantum states are considered to evolve along and against the flow of time \cite{Laflamme}.
We note that possible paradoxes with causality are avoided because the necessity of additional classical communications~\cite{Hardy}.
These effects have been actively studied by using postselection in the frameworks of ``conditional time travel'' \cite{Laflamme} and projective closed timelike curves \cite{Lloyd}.

We note that, in the standard one-qubit teleportation scheme, the direction of an established channel is governed by operations performed by two parties [see Fig.~\ref{fig:F0}(a)].
Indeed, the state is transferred from the party conducting a projective measurement (Alice) to the party applying an unitary transformation (Bob).

In our work, 
we address a question about the possibility to modify operations performed by parties in order to adapt the information-transferring medium of Bell state for carrying of two qubits spreading in opposite 
directions: from Alice to Bob, and from Bob to Alice [see Fig.~\ref{fig:F0}(b)].
This study of bidirectional (two-way) teleportation (together with the first protocol for continuous variables teleportation) has been initiated by the scheme in Ref.~\cite{Vaidman}.
A direct way for bidirectional quantum teleportation is to use two quantum teleportations with two Bell states: from Alice to Bob and vice versa.
A number of alternative approaches to bidirectional quantum teleportation based on $n$-qubit entangled states with $n>4$ \cite{Xin,Yan,Duan2,Thapliyal} and GHZ states \cite{Hassanpour} 
has been suggested. 

Our approach is different from that mentioned above in two points. 
First, our protocol occupies the minimal amount of shared entangled qubits, {\it i.e.}, a singe Bell state pair.
Second, 
the fee for such utilization of shared entanglement is that bidirectional transmission of quantum states becomes imperfect and only noisy versions of input states can be obtained at the output.
Bidirectional teleportation with a Single Bell state can be interesting both from the fundamental viewpoint, since it uncovers the additional potentialities of shared entanglement, 
and from practical perspectives because experimental control for Bell states is on a higher level than for many-qubit states.

We study three schemes for implementation of bidirectional quantum teleportation with a single Bell state and auxiliary two-level quantum systems (trigger qubits), 
whose initial states control the workflow of the schemes.
First, we consider a scheme, where the actions of parties are governed by two independent random trigger qubits.
In the second scheme, a single random trigger is used.
The third one is a mixture between first and second schemes and can be considered as having tunable level of classical correlation between trigger qubits on both sides.

An important benchmark for imperfect quantum teleportation to ensure that the resource of quantum entanglement is utilized is to require the following inequality for the teleportation fidelity:
$F>F_{\rm class}$, where $F_{\rm class}=2/3$ is the maximal fidelity that can be achieved by means of classical communication only~\cite{Oh}. 
We analyze the fidelity of teleportation for all schemes and find a minimal amount of correlation between trigger qubits, 
which allows the classical fidelity boundary to be overcome 
(as will be shown the scheme with independent trigger qubits fails to simultaneously achieve $F_{\rm class}$ in both directions).

The fact that appearing quantum channels belong to the class of depolarizing ones also allows us to employ the Choi-Jamiolkowski isomorphism~\cite{Jamiolkowski,HolevoBook} 
to study how classical capacity~\cite{Wootters2}, entanglement-assisted capacity~\cite{Thapliyal2}, 
and the entanglement breaking property~\cite{Holevo,HolevoBook,Horodecki} 
of the suggested teleportation channels relate to the auxiliary classical information responsible for forming correlations between trigger qubits.

The paper is organized as follows:
We suggest bidirectional teleportation schemes with two independent quantum random trigger qubits, 
one random trigger qubit, 
and a mixture between the previous schemes in Sec. \ref{sec:scheme}.
In Sec. \ref{sec:fidelity}, we calculate the fidelity of teleportation for all schemes and find a condition on correlation between trigger qubits in the mixed scheme, 
which allows us to overcome the classical fidelity boundary $F_{\rm class}$.
We analyze the relation between characteristics of Choi-Jamiolkowski state and the amount of auxiliary classical communication in Sec. \ref{sec:entanglement}.
Finally, our results are summarized in Sec. \ref{sec:conclusion}.

\section{Bidirectional teleportation}\label{sec:scheme}

\begin{figure*}[htbp]
\includegraphics[width=1\linewidth]{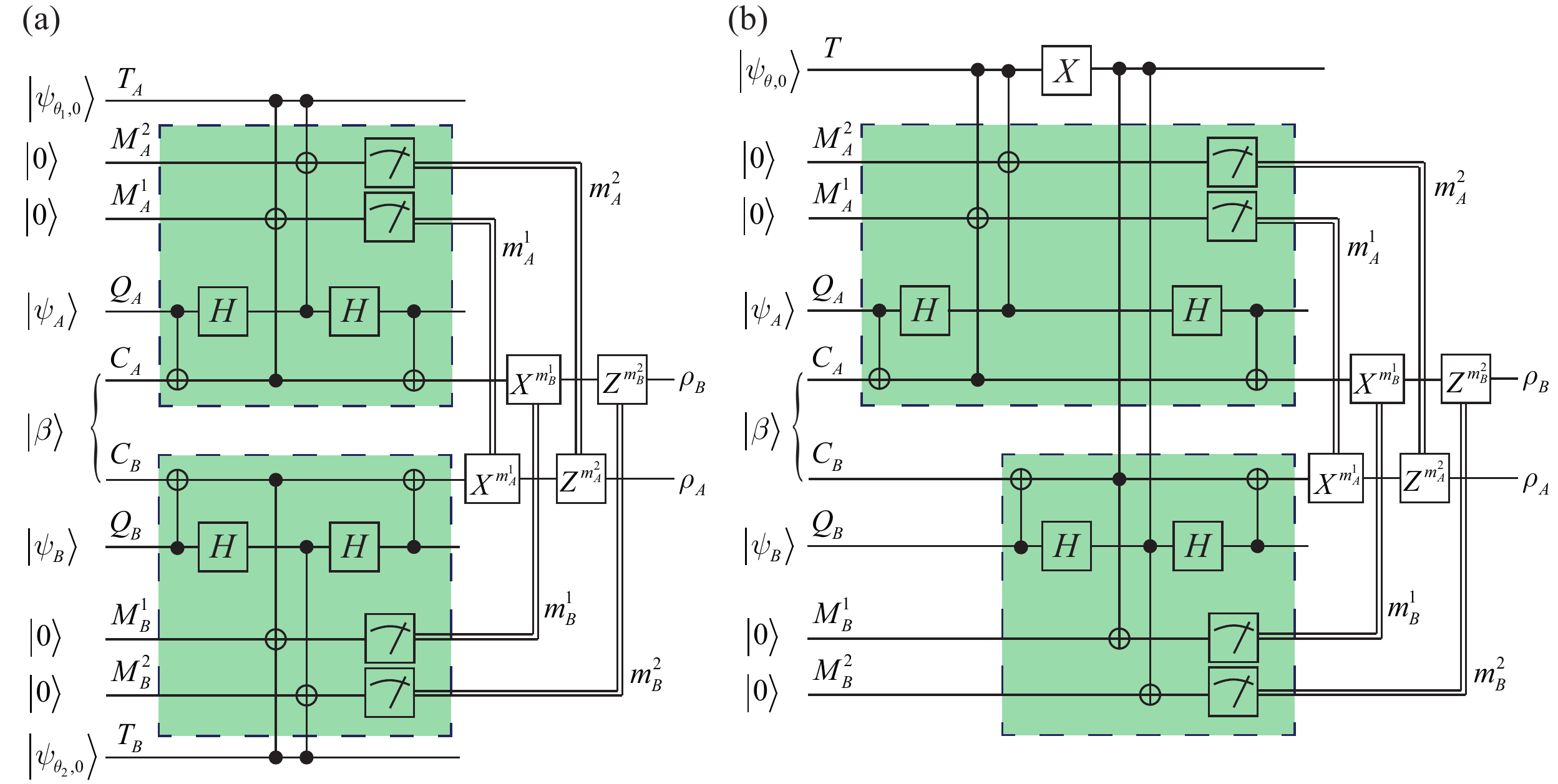}
\vskip -4mm
\caption{Quantum circuits for implementation of suggested schemes for bidirectional teleportation with the single Bell state $\ket{\beta}$ [See Eq. (\ref{eq:BellState})].
	(a) Scheme with using two independent qubit triggers. 
	(b) Scheme with a single qubit trigger.
	Here $X$, $Z$ and $H$ are Pauli $\sigma_x$, $\sigma_z$ and Hadamard gates respectively.
	Vertical lines represent CNOT and CCNOT (Toffoli) gates, where control and target qubits are denoted by $\bullet$ and $\oplus$ correspondingly.
	All measurements are performed in the computational basis with outcomes $0$ or $1$.
	The observed values ($m^i_\alpha$) are transmitted via classical communication represented by thick lines toward gates, which are implemented whether the value is 1.	
	The highlighted parts (in dashed-line blocks) of the circuits serve as indirect Bell measurements controlled by trigger qubits.} 
\label{fig:TelSchemes}
\end{figure*}

The standard one-qubit quantum teleportation protocol works as follows.
The idea is to transfer the state $\ket{\psi}$ from one party (Alice) to another (Bob).
Before protocol start, Alice and Bob share two particles of a maximally entangled pair.
We consider one of the Bell states
\begin{equation} \label{eq:BellState}
	\ket{\beta}=\frac{1}{\sqrt{2}}(\ket{00}+\ket{11}).
\end{equation}
The protocol consists of three steps: 
(i) Alice performs a projective measurement in the Bell basis of her particle in the state $\ket{\psi}$ and the particle from the entangled pair; 
(ii) Alice sends the index of the measurement outcome (that is 2 bits of information) to Bob through an ideal classical communication channel; 
(iii) Bob performs the particular unitary operation, which depends on message from Alice, on his particle from the entangled pair to obtain it in the state $\ket{\psi}$.

We note that Alice's measurement in the first step of the protocol breaks an initial entanglement of state (\ref{eq:BellState}), 
and her particle from the entangled pair transforms to a maximally mixed state.
Besides, Bob has to minimize an interaction of his particle from the entangled pair with an environment to avoid decoherence that reduces the fidelity of the quantum information transfer.

These two points indicate that, to make it possible to transfer information from Bob to Alice as well as from Alice to Bob, 
{\it i.e.}, make teleportation bidirectional, Alice should not perform a perfect projective measurement of her particles.
Indeed, Alice should leave some correlations of the Bell state for the purpose of information transfer in the reverse direction.
Bob should be guided by the same strategy to obtain the quantum state from Alice and, at the same time, send his own one.

Below, we describe possible schemes for bidirectional quantum teleportation with two random trigger qubits, single random trigger qubits, and a mixture between these two schemes. 
The suggested method of entanglement utilization results in imperfections of the transmitted states.
In other words, teleportation fidelities are less than unity.
Nevertheless, in such a scheme the fidelity of teleportation can be high enough for applications in quantum information technologies. 
We also note that the scheme operates with $N=10$ qubits; however, these qubits should be prepared locally for two parties without need to be transferred. 
This looks realistic in the view of recent experimental advances in operating with multiqubit systems \cite{Blatt2}. 

\subsection{Independent random triggers}

The idea of the first scheme is based on two independent random trigger qubits that guide the actions of Alice and Bob.
The scheme  operates with $N=10$ qubits as it is presented in Fig.~\ref{fig:TelSchemes}a.
Here qubits with subindices $A$ and $B$ belong to Alice and Bob correspondingly.
The qubits have the following assignments:
(i) $C_{A}$ ($C_B$) are initialized with the Bell state $\ket{\beta}$~(\ref{eq:BellState}) and they are used for the transmission of quantum information;
(ii) $Q_{A}$ ($Q_{B}$) are initialized with the states $\ket{\psi_{A}}$ ($\ket{\psi_{B}}$), which Alice (Bob) wants to transmit;
(iii) $M_{A}^{1}$ ($M_{B}^{1}$) and $M_{A}^{2}$ ($M_{B}^{2}$) are initialized with the state $\ket{0}$ and they are used for storage
 of projective measurement outcomes;
(iv) $T_{A}$ ($T_B$) are trigger qubits initialized with the states $\ket{\psi_{\theta_{1},0}}$ ($\ket{\psi_{\theta_{2},0}}$), 
where
\begin{equation} \label{eq:BlSpher}
	\ket{\psi_{\theta,\phi}}=\cos\frac{\theta}{2}\ket{0}+e^{i\phi}\sin\frac{\theta}{2}\ket{1}
\end{equation}
is the standard parametrization of a qubit pure state on the Bloch sphere.

To perform the projective measurement in the Bell basis using measurements in computational basis, 
one can use a sequence of controlled-NOT (CNOT) 
and Hadamard gates that converts a set of four Bell state vectors into a set of four two-qubit computational basis vectors.
This sequence is applied to qubits $Q_{A}$ ($Q_{B}$) and $C_{A}$ ($C_B$).
Next Toffoli (CCNOT) gates make `copies' (in the sense of computational basis) of qubits 
$C_{A}$ ($C_{B}$) and $Q_{A}$ ($Q_{B}$) on $M_{A}^1$ ($M_{B}^1$) and $M_{A}^2$ ($M_{B}^2$) if the triggers $T_{A}$ ($T_{B}$) are in the state $\ket{1}$.
Then the sequences of $\mathrm{CNOT}$ and Hadamard gates are repeated in reverse order to 
undo the change of basis for $C_{A}$ ($C_{B}$) and $Q_{A}$ ($Q_{B}$).
$M_{A}^{1}$ ($M_{B}^{1}$) and $M_{A}^{2}$ ($M_{B}^{2}$)  are measured in the computational basis, and the corresponding results $m_{A}^{1}$ ($m_{B}^{1}$) and $m_{A}^{2}$ ($m_{B}^{2}$) are sent to the opposite party.
On the opposite side, the appropriate unitary transformations as in standard teleportation protocol are applied.
The teleported states of Alice (Bob) are obtained on qubits $C_{B(A)}$. 

The resulting quantum channels from Alice to Bob and from Bob to Alice can be written as
\begin{equation} \label{eq:EInd}
\begin{aligned}
	\EindAB[\rhoin]=p_1(1-p_2)\rhoin+\left[1-p_1(1-p_2)\right]\rhom, \\
	\EindBA[\rhoin]=p_2(1-p_1)\rhoin+\left[1-p_2(1-p_1)\right]\rhom,
\end{aligned}
\end{equation}
where $p_i=\sin^2(\theta_i/2)$ and $\rho_0$ is the maximally mixed qubit state. 

Let us consider several important cases.
If $\theta_1=\theta_2=0$, then neither Alice nor Bob make projective measurements, 
the entangled state of $\ket{\beta}$ of $C_AC_B$ remains unchanged, and the parties end up with maximally mixed state $\rho_A=\rho_B=\rho_0$.
If $\theta_1=\pi$ and $\theta_2=0$, then Alice performs a perfect projective measurement, 
Bob performs an appropriate unitary transformation and obtains his particle $C_B$ in the desired state $\rho_A=\ket{\psi_A}\bra{\psi_A}$, while Alice ends up with $\rho_B=\rho_0$.
That is perfect unidirectional teleportation from Alice to Bob.
In the case of $\theta_1=0$ and $\theta_2=\pi$, 
we obtain perfect unidirectional teleportation from Bob to Alice.
If $\theta_1=\theta_2=\pi$, 
then both parties make perfect projective measurements on their particles and obtain maximally mixed states $\rho_A=\rho_B=\rho_0$ in the result.

\subsection{Common random trigger}
The second scheme [see Fig.\ref{fig:TelSchemes}b] looks similar to that considered above;
however, here we exchange two independent trigger qubits $T_A$ and $T_B$ with a single common random trigger qubit $T$ initialized with the state  $\ket{\psi_{\theta,0}}$.
The trigger qubit guides the performing of projective measurements on both sides simultaneously in such a way that it happens only on the one side.
This is achieved by applying the inverse gate $X$ on the trigger qubit $T$.  
For this scheme, the two channels are as follows:
\begin{equation} \label{eq:ECom}
	\begin{aligned}
		\EcomAB[\rhoin]=p\rhoin+(1-p)\rhom, \\
		\EcomBA[\rhoin]=(1-p)\rhoin+p\rhom,
	\end{aligned}
\end{equation}
where $p=\sin^2(\theta/2)$.
In this way, we obtain teleportation from Alice (Bob) to Bob (Alice) at $\theta=\pi(0)$.

There are two points about common trigger qubits we should mention.
(i) Although its control over Alice's and Bob's indirect Bell measurements [see highlighted boxes in Fig.~\ref{fig:TelSchemes}(b)]  are presented by CCNOT gates, 
the control could be performed via classical communications as well: the measured value of $T$'s initial state is used by Alice, and its inverted value is sent to Bob. 
(ii) Trigger qubit $T$ can belong to the third independent party (Charlie) and its random value can be transmitted via a classical channel to Alice and Bob.

\subsection{Mixed scheme}
Finally, one can consider a mixture of the two schemes, governed by parameter $t\in[0,1]$:
\begin{equation} \label{eq:EMix}
	\begin{aligned}
		\EmixAB[\rhoin]=t\EindAB[\rhoin]+(1-t)\EcomAB[\rhoin],\\
		\EmixBA[\rhoin]=t\EindBA[\rhoin]+(1-t)\EcomBA[\rhoin],
	\end{aligned}
\end{equation}
where at $t=0$ and $t=1$ we obtain the schemes with common and independent triggers correspondingly.
This scheme can be considered as intermediate between two limiting situation: 
the first one with two parties acting independently from each other, and the second one with the two parties acting in complete correlation to obtain a constructive result.
In this way, the mixed scheme employs tunable level of classical correlation between trigger qubits on both sides. 

\section{Fidelity of teleportation}\label{sec:fidelity}

The suggested scheme for bidirectional quantum teleportation via a single Bell state has an important  shortcoming: it can not be presented by a perfectly identical channel.
Then the scheme can be mapped on the conventional unidirectional teleportation with the use of not maximally entangled states.

For a qubit quantum channel $\mathcal{E}$, 
the standard measure of imperfections is the averaged fidelity for all pure input state vectors over the Bloch sphere:
\begin{equation} \label{eq:Fid}
	F\left[\mathcal{E}\right]=\frac{1}{4\pi}\int_{0}^{\pi}d\theta\int_{0}^{2\pi}d\phi f(\mathcal{E};\theta,\phi)\sin\theta,
\end{equation}
where 
\begin{equation}
	f(\mathcal{E};\theta,\phi)=\bra{\psitp}\mathcal{E}\left[\ket{\psitp}\bra{\psitp}\right]\ket{\psitp}.
\end{equation}
For the teleportation protocol the expression~(\ref{eq:Fid}) is known as fidelity of teleportation.
In view of the critical value of $F_{\rm class}=2/3$, which can be obtained by using only classical communication \cite{Oh}, 
we are interested in the regimes, where teleportation channels demonstrate the averaged fidelity larger than $2/3$.

\begin{figure}[t]
	\includegraphics[width=1\linewidth]{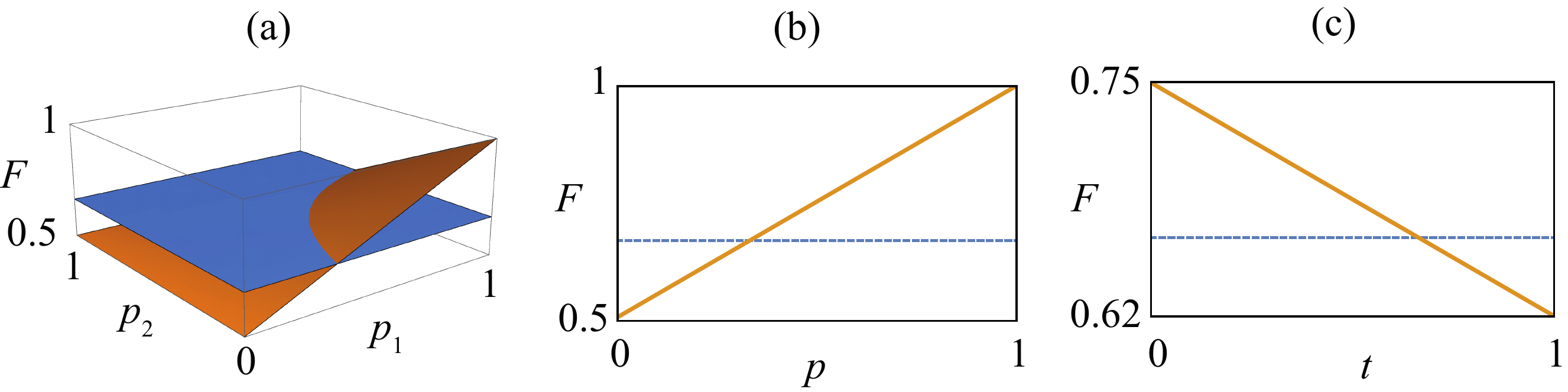}
	\vskip -4mm
	\caption{Fidelities of teleportation from Alice to Bob for different modifications of bidirectional teleportation.
	Fidelity of teleportation given by Eq. (\ref{eq:EFid}) is shown:
	(a) for the scheme with independent trigger qubits as function of $p_1$ and $p_2$;
	(b) for the scheme with a single common trigger qubit as function of $p$, and
	(c) for the mixed scheme with $p=p_1=p_2=1/2$ as a function of $t$.
	The horizontal dashed lines (plane) correspond to the critical value $F_{\rm class}=2/3$.}
	\label{fig:Fid}
\end{figure}

By substituting (\ref{eq:EInd})--(\ref{eq:EMix}) into (\ref{eq:Fid}), 
for the considered bidirectional teleportation schemes [see Fig.~\ref{fig:Fid}] we have:
\begin{equation} \label{eq:EFid}
\begin{aligned}
	&F[\EindAB]=\frac{1}{2}\left[1+p_1(1-p_2)\right], \\
	&F[\EcomAB]=\frac{1}{2}(1+p), \\
	&F[\EmixAB]=\frac{1}{2}\left\{1+p+t\left[p_1(1-p_2)-p\right]\right\}
\end{aligned}
\end{equation}
We note that fidelities for the inverse direction $B\rightarrow A$ can be obtained be exchange $p_1\leftrightarrow p_2$ and $p\leftrightarrow 1-p$.

In the symmetrical regime $p_1=p_2=p=1/2$ for the schemes with independent and common trigger qubits, 
we obtain the averaged fidelities $F[\Eind]=0.625<2/3$ and $F[\Ecom]=0.75>2/3$.
Then we conclude that the scheme with independent triggers is inefficient. 
Better results can indeed be obtained just by direct measurements and classical communication.
Nevertheless, for the common trigger scheme the fidelity is high enough to observe benefits of using a shared entangled state.

In the case of the mixed scheme in the symmetric regime we obtain $F[\Emix]=3/4-t/8$, which demonstrates linear interpolation between limiting cases considered.
The solution for the critical level $F[\Emix]=2/3$ is $t_0=2/3$.
This value corresponds to a minimal amount of correlation between the actions of Alice and Bob,
which allows us to approach the quantum regime in the symmetric bidirectional teleportation scheme.

\section{Information flows analysis}\label{sec:entanglement}

We consider information transfer in the proposed bidirectional version of quantum teleportation protocol.
We focus on the mixed scheme in the symmetric operating mode, where Alice and Bob transmit equal amounts of information to each other.
The resulting quantum channels (same in both directions) are given by the expression
\begin{equation}\label{eq:sr}
	\Emix[\rhoin]=\left(\frac{1}{2}-\frac{t}{4}\right)\rhoin+\left(\frac{1}{2}+\frac{t}{4}\right)\rhom,
\end{equation}
which is obtained by substituting $p_1=p_2=p=1/2$ in Eq.~\eqref{eq:EMix}.

We start from consideration of auxiliary classical communication that is needed for establishing correlations between operations of the parties.
For that purpose we introduce a matrix of joint probability distribution
\begin{equation}\label{eq:Probs}
	\mathcal{M}=\frac{t}{4}\begin{bmatrix}
	1 & 1 \\
	1 & 1
	\end{bmatrix}+\frac{1-t}{2}
	\begin{bmatrix}
	0 & 1 \\
	1 & 0
	\end{bmatrix},
\end{equation}
whose rows (columns) correspond to the decision of making a projective measurement by Alice (Bob).
The first and second terms in Eq.~(\ref{eq:Probs}) are related to independent and common trigger qubits correspondingly.

From the joint distribution (\ref{eq:Probs}), one can extract the marginal distributions
\begin{equation}
	\mathcal{M}^A_i=\sum_j \mathcal{M}_{ij}, \quad \mathcal{M}^B_j=\sum_i \mathcal{M}_{ij},
\end{equation}
which in turn allows us to calculate the Shannon's mutual information
\begin{equation} \label{eq:ShInf}
	\mathcal{I}_\mathrm{aux} = \sum_{ij}\mathcal{M}_{ij}\log_2\frac{\mathcal{M}_{ij}}{\mathcal{M}^A_{i}\mathcal{M}^B_{j}}.
\end{equation}
The value of $\mathcal{I}_\mathrm{aux}$ gives the auxiliary classical communication, and for the symmetric mixed scheme under consideration is given by 
\begin{equation}\label{eq:clinf}
	\mathcal{I}_\mathrm{aux}=2-h_4^{2,2}\left(t/4\right),
\end{equation}
where
\begin{equation}
h_4^{2,2}(x)=-2x\log_2x-(1-2x)\log_2(1/2-x)
\end{equation}
is a form of the one-parameter quaternary entropy function. 

Obviously, we obtain $\mathcal{I}_\mathrm{aux}=1$ for $t=0$ (common trigger scheme) and $\mathcal{I}_\mathrm{aux}=0$ for $t=1$ (independent triggers scheme). 
For the critical value $t=t_0=2/3$, 
we have the $\mathcal{I}_\mathrm{aux}=\mathcal{I}_0\approx0.0817$ bits, 
which turns out to be the minimal amount of auxiliary classical communication between parties necessary to overcome a nonclassical teleportation fidelity.

\begin{figure}[t]
	\includegraphics[width=0.95\linewidth]{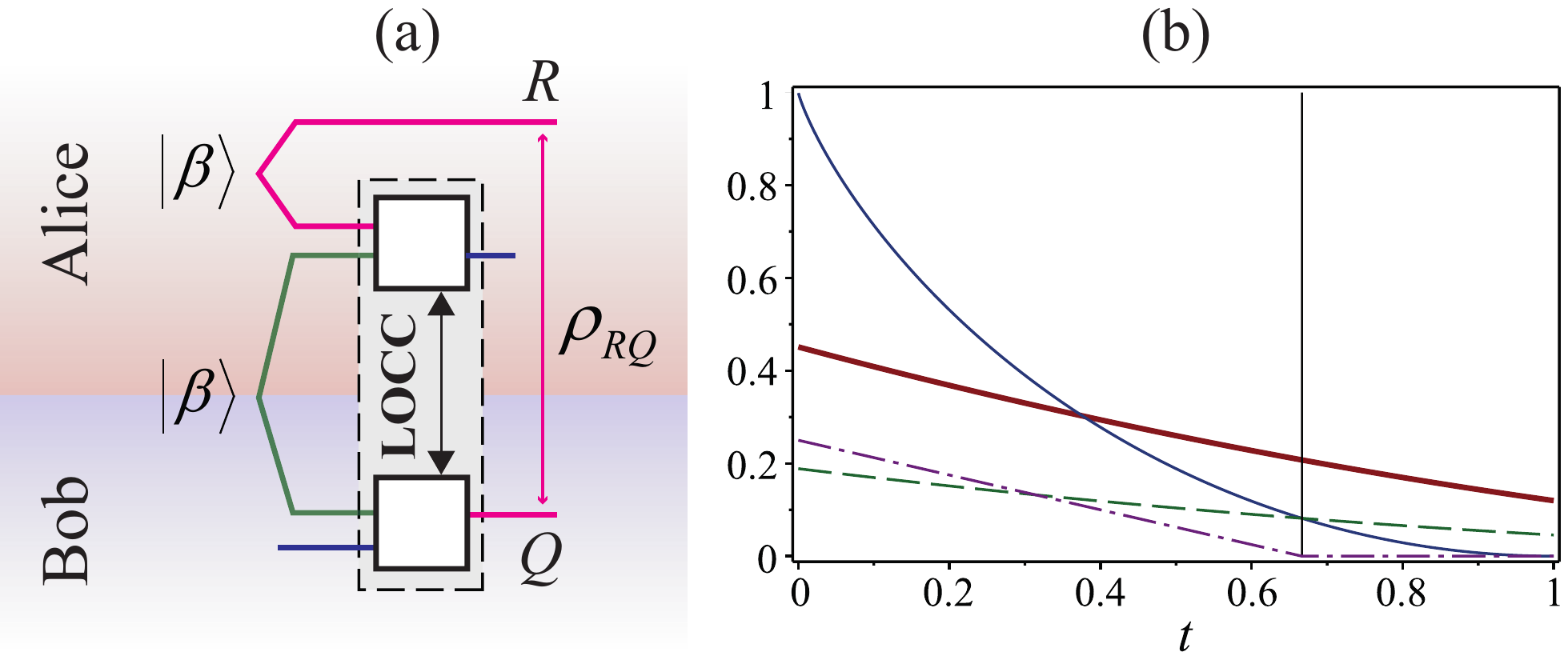}
	\vskip -4mm
	\caption{
	(a) Representation of the Choi-Jamiolkowski state $\rho_{RQ}$, which is used for investigation of information flows in the symmetric bidirectional teleportation.
	(b) Different measures of information flows in the mixed scheme of symmetric bidirectional teleportation as a function of parameter $t$: 
	auxiliary classical information $\mathcal{I}_{\rm aux}$ (thin solid curve); 
	entanglement-assistant capacity $\mathcal{I}_{\rm tot}$ (thick solid curve); 
	classical capacity $\mathcal{I}_{\rm class}$ (dashed curve); 
	concurrence $\mathcal{C}$ of the state $\rho_{RQ}$ (dash dot curve).
	Vertical line corresponds to the critical value $t_0=2/3$.}
	\label{fig:Infs}
\end{figure}

We also analyze information transfer directly through quantum channel (\ref{eq:sr}).
Along this line, we imply the Choi-Jamiolkowski isomorphism~\cite{Jamiolkowski,HolevoBook}, 
which establishes a compliance between quantum channel and quantum state.
We consider the Choi-Jamiolkowski state $\rho_{RQ}$, 
that is obtained by propagation of a part of a maximally entangled state~(\ref{eq:BellState}) through a teleportation channel, 
{\it e.g.}, from Alice to Bob [see Fig.~\ref{fig:Infs}].
The passing qubit is denoted as $Q$, the untouched (reference) qubit as $R$, 
and the resulting density matrix is given by
\begin{equation}\label{eq:chst}
	\rho_{RQ}=\left(\frac{1}{2}-\frac{t}{4}\right)\ket{\beta}\bra{\beta}+\left(\frac{1}{2}+\frac{t}{4}\right)\rhom\otimes\rhom.
\end{equation}

As a measure of total information transfer through a quantum channel~(\ref{eq:sr}), 
we use the following form of the quantum mutual information:
\begin{equation}\label{eq:QMI}
	\mathcal{I}_\mathrm{tot} = S[\rho_R]+S[\rho_{Q}]-S[\rho_{RQ}],
\end{equation}
where $S[\rho]=-\mathrm{Tr}[\rho\log_2\rho]$ is the von Neumann entropy and $\rho_R$ and $\rho_{Q}$ are the reduced states of $\rho_{RQ}$.
We note that $\mathcal{I}_\mathrm{tot}$ also gives an entanglement-assisted capacity of our depolarizing channel~\cite{HolevoBook}.
By substituting (\ref{eq:chst}) into Eq. (\ref{eq:QMI}), we obtain 
\begin{equation}\label{eq:qinf}
	\mathcal{I}_\mathrm{tot} = 2-h_4^{3,1}\left({1}/{8}+{t}/{16}\right),
\end{equation}
with another version of the one-parameter quaternary entropy function
\begin{equation}
	h_4^{3,1}(x)=-3x\log_2x-(1-3x)\log_2(1-3x).
\end{equation}

The behavior of the auxiliary classical communication $\mathcal{I}_\mathrm{aux}$ 
and quantum mutual information (entanglement-assisted capacity) $\mathcal{I}_\mathrm{tot}$ is depicted by solid lines in Fig.~\ref{fig:Infs}.
While the additional classical communication $\mathcal{I}_\mathrm{aux}$ decreases from 1 to 0, the quantum mutual information decreases from $0.451$ to $0.120$.
In this way additional classical communication increases a throughput of resulting depolorizing channels more than in 3.5 times.

To split classical and quantum components of quantum mutual information~(\ref{eq:qinf}), 
we consider the quantum discord~\cite{Discord} of the Choi--Jamiolkowski state, which is given by 
\begin{equation}
	\mathcal{D}=\mathcal{I}_\mathrm{tot}-\mathcal{I}_\mathrm{class}
\end{equation}
with the classically-accessible information \cite{Zurek}
\begin{equation}
	\mathcal{I}_\mathrm{class} = 
	\max_{\{\Pi^R_j\}}\left(
	S[\rho_Q]-\sum_j p_{\Pi_j^R}S[\rho_{Q|\Pi^R_j}]
	\right)
\end{equation}
where $\{\Pi^R_j\}$ is the positive-operator valued measure (POVM) in the space of reference qubit $R$, 
\begin{equation}
	p_{\Pi_j^R}=\Tr[\Pi_j^R\rho_R]
\end{equation}
is a probability of $j$-th measurement outcome, 
and 
\begin{equation}
	\rho_{Q|\Pi_j^R}=p_{\Pi_j^R}^{-1}\Tr_R[\rho_{RQ}\Pi_j^R]
\end{equation}
is a corresponding conditional state of $Q$.
For the considered state~\eqref{eq:chst} the optimal measurement turns out to be any von Neumann measurement, and the value of classically-accessible information is given by
\begin{equation} \label{eq:clcap}
	\mathcal{I}_\mathrm{class} =1-h_2\left(3/4-t/8\right),
\end{equation}
where
\begin{equation}
	h_2(x)=-x\log_2x-(1-x)\log_2(1-x)
\end{equation}
is a binary entropy function.
It is remarkable that the value of $\mathcal{I}_\mathrm{class}$ also gives the classical capacity of the channel~\cite{HolevoBook}.
It follows from the fact that the classical capacity of the depolarizing channel is achieved on sets of two orthogonal states 
input into the channel with equal probabilities.
At the same time the von Neumann measurement of one particle from a maximally entangled pair is also considered as remote preparation (although not controlled) of such the input.  

The behavior of classically-accessible information (classical capacity) $\mathcal{I}_\mathrm{class}$ as a function of parameter $t$ is given by dashed line in~Fig.\ref{fig:Infs}.
It decreases from 0.189 bits at $t=0$ to 0.0456 at $t=1$, so the auxiliary communication increases classical capacity more than in 4.1 times.
It is quite remarkable that the curve of $\mathcal{I}_\mathrm{class}$ crosses the curve of $\mathcal{I}_\mathrm{aux}$ exactly at the critical point $t=t_0$.
At $t<t_0$ an overcoming of classical auxiliary communication over classical capacities turns the teleportation channel into a ``quantum regime'' with $F>F_{\rm class}$.
Probably this may be due to the possibility of transferring this auxiliary information via one of the teleportation channels.
Nevertheless, we leave a deeper investigation of this identity and its possible generalization on asymmetric regime, where the capacities of opposite channels are different, for further study.

Another important characteristic that describes nonclassical behavior of a quantum channel is its ability to maintain the initial entanglement between a passing system 
and some external system that does not interact with the channel and its local environment.
The channels, whose action on one part of the entangled states makes these state separable, are known as entanglement breaking channels~\cite{Holevo}.
The criterion for entanglement breaking for a particular channel is separability of the corresponding Choi-Jamiolkowski state~\cite{Horodecki}.
To investigate this property we calculate the concurrence, which is a measure of entanglement in two-qubit states~\cite{Conc}.
It is given by 
\begin{equation}\label{eq:Conc}
	\mathcal{C}=\max{\left(0,{\lambda}_1-{\lambda}_2-{\lambda}_3-{\lambda}_4\right)},
\end{equation}
where $\{{\lambda}_i\}$ is a set of eigenvalues (in descending order) of the density matrix
\begin{equation}	
\widetilde{\rho}_{RQ}=\sqrt{\sqrt{\rho_{RQ}}(\sigma_y\otimes\sigma_y)\rho_{RQ}^{*}(\sigma_y\otimes\sigma_y)\sqrt{\rho_{RQ}}},
\end{equation}
with $\sigma_y$ being the standard Pauli matrix and $*$ stands for complex conjugation.

For the particular density matrix of the form (\ref{eq:chst}), we obtain
\begin{equation}\label{eq:concval}
	\mathcal{C}=\max\left(0,1/4-3t/8\right).
\end{equation}
As we see perfect correlation between operations of the parties ($t=0$) results in one quarter of the maximum value (dashed line in Fig.~\ref{fig:Infs}).
One can also see that the channels become entanglement breaking at the point $t=t_0$, where the fidelity of teleportation reaches the classical boundary.
This behavior can be explained by the fact that any entanglement breaking channel can be expressed 
as a measure-and-prepare channel which consists of intermediate stage where all information is encoded as a classical state~\cite{Horodecki}.
In our case, this stage corresponds to classical communication between parties.

Finally, we would note the coherent information responsible for quantum capacity~\cite{Lloyd2} and given by
\begin{equation}
	\mathcal{I}_\mathrm{coh}=S[\rho_Q]-S[\rho_{RQ}]
\end{equation}
is negative for the whole region of parameter $t$.
It can be easily revealed using the fact that the reduced state $\rho_R$ is maximally mixed, therefore
\begin{equation}
	\mathcal{I}_\mathrm{coh}=\mathcal{I}_{\rm tot}-1,
\end{equation}
while the value of $\mathcal{I}_{\rm tot}$ is less than 1/2 [see Fig.~\ref{fig:Infs}(b)].

\section{Conclusions and outlook}\label{sec:conclusion}

We now summarize the main results of the present paper. 
We suggested several approaches to implement a bidirectional version of the standard one-qubit teleportation protocol with the use of a single Bell state:
with two independent quantum random trigger qubits, 
with one random trigger qubit,
and mixture between the first and second approaches.
We demonstrated that it is possible to achieve a regime of overcoming the classical fidelity boundary of $2/3$ for teleportation channels in both directions, and showed how the auxiliary classical communication, using for establishing correlations between trigger qubits, amplifies the capacities of the resulting quantum channels.
We also revealed that the operating mode, 
where the fidelities of teleportation reach the classical boundary, corresponds to the identity between auxiliary classical communication and classical capacity of the channels.

Current state of the art demonstrates a potential for realization of the suggested bidirectional imperfect quantum teleportation with a single Bell state 
and a necessity of further investigations of the suggested scheme, {\it e.g.}, in the framework of the logic Bell-state analysis \cite{Sheng}.

\section*{Acknowledgements}

The authors thank the anonymous referees for useful comments. 
The support from Ministry of Education and Science of the Russian Federation in the framework of the Federal Program (Agreement 14.579.21.0104) is acknowledged.

\end{document}